\documentclass[prd,nofootinbib,english,twocolumn]{revtex4-2}
\usepackage[utf8]{inputenc}

\usepackage{graphicx,color} 
\usepackage{xcolor}
\usepackage{rotating}
\usepackage{amssymb}
\usepackage{amsfonts}
\usepackage{amsmath}
\usepackage{xspace}
\usepackage{mathtools} 
\usepackage{verbatim}
\usepackage{comment}
\usepackage{enumitem}
\usepackage{physics}
\usepackage{graphicx}
\usepackage[colorlinks]{hyperref}
\usepackage{MnSymbol}
\usepackage{graphicx}
\usepackage{accents}

\renewcommand{\vec}[1]{\boldsymbol{#1}}

\begin{document}

\title{ 
Gravitational attraction of ultra-relativistic matter: A new testbed for modified gravity at the Large Hadron Collider
}

\author{Christian Pfeifer}
\email{christian.pfeifer@zarm.uni-bremen.de}
\affiliation{ZARM, University of Bremen, 28359 Bremen, Germany.}

\author{Dennis R\"atzel}
\email{dennis.raetzel@zarm.uni-bremen.de}
\affiliation{ZARM, University of Bremen, 28359 Bremen, Germany.}

\author{Daniel Braun }
\email{daniel.braun@uni-tuebingen.de}
\affiliation{Institut f\"{u}r Theoretische Physik, Eberhard-Karls-Universit\"{a}t T\"{u}bingen, 72076 T\"{u}bingen, Germany}

\begin{abstract}
We derive the scalar-tensor modification of the gravitational field of an ultrarelativistic particle beam and its effect on a test particle that is used as sensor. To do so, we solve the linearized scalar-tensor gravity field equations sourced by an energy-momentum tensor of a moving point particle. The geodesic equation and the geodesic deviation equation then predict the acceleration of the test particle as well as the momentum transfer due to a passing source. Comparing the momentum transfer predicted by general relativity and scalar tensor gravity, we find that there exists a relevant parameter regime where this difference increases significantly with the velocity of the source particle. Since ultrarelativistic particles are available at accelerators like the Large Hadron Collider, ultra-precise acceleration sensors in the vicinity of the particle beam could potentially detect deviations from general relativity or constrain modified gravity models.
\end{abstract}

\maketitle

\section{Introduction}
Our understanding and description of the gravitational interaction based on General Relativity (GR) and the standard model of particle physics, is still very much incomplete. A fact, that is easily realized by the necessity to conclude that the constituents of the universe are by large amount dark matter ($\sim 27\% $) and dark energy ($\sim 68\%$), in order to explain dynamics of galaxies and the universe as a whole \cite{Planck:2013oqw}. In cosmology, even this $\Lambda CDM$ model reaches its limits, as the tensions on the early time and late time observation of the Hubble constant $H_0$ and density fluctuations $S_8$ (or $\sigma_8$) demonstrate \cite{Abdalla:2022yfr}. Moreover, infinite gravitational tidal forces and singularities are inevitable predictions of GR and an understanding of gravity in terms of a quantum field theory is still elusive \cite{Addazi:2021xuf}.

To improve our understanding of the gravitational force numerous frameworks, models and theories have been suggested to extend GR or the standard model of particle physics or both. Up to date, none of them could explain satisfactorily all discrepancies between theory and observation or the theoretical difficulties in our description of gravity.

To scrutinize extension of GR, the first applications and tests are usually performed in the context of cosmology~\cite{CANTATA:2021ktz}, by the study of compact objects \cite{Casado-Turrion:2023jba} or in a post-Newtonian weak field and slow velocity expansion \cite{damour1996tensor,Freire:relativ2012,will2014confrontation}, meaning in highly symmetric situations or in the non-relativistic regime. 

Rarely, the impact of extensions of GR is investigated for very fast ultrarelativistic sources of gravity \cite{GERTSENSHTEIN}. Physically, such sources exist for example in violent astrophysical environments in the stage shortly before merging in inspirals of binary systems and leave a trace in gravitational wave signals \cite{Berti:2018cxi}.

Another physical setup, which offers an opportunity where the gravitational field of ultrarelativistic sources might be detected, are particle accelerators. The proton bunches at the Large Hadron Collider (LHC) generate a gravitational field that is mainly sourced by the kinetic energy and momentum of the particles, compared to which their rest mass is negligible. Thus sensitive detectors placed around the beam line of the LHC could be able to test the behaviour of the gravitational interaction, and the corresponding theoretical models, in  a very controlled environment \cite{Spengler:2021rlg}, in contrast and complementary to the astrophysical systems. 

In this article we propose to analyse and test extensions and modifications of GR in the ultrarelativistic limit. This means we consider gravitating objects, moving close to the speed of light which produce a weak gravitational field, so that one can employ linearized gravity. The goal is to predict the gravitational acceleration of a test mass near an ultrarelativistic particle beam in a particle accelerator from modified theories of gravity, in order to test their predictions complementary on the one hand to the cosmological, or astrophysical, strong field regime and on the other hand to the post-Newtonian slow-velocity regime. In order to demonstrate the principle and the effect, we consider a wide class of Brans-Dicke-inspired parametrized scalar-tensor theories, which has been analyzed in its post Newtonian regime in all generality \cite{Hohmann:2013rba}. Among others, the theories we investigate include Brans-Dicke theory itself and the scalar tensor theory representation of $f(R)$ theories \cite{Teyssandier:1983zz,Faulkner:2006ub}. In a complementary work \cite{Damour:1995kt}, the post-Minkowskian interaction of two gravitating particles, i.e.\ not assuming that the second particle is a test particle, has been discussed. We focus on one particle being a test particle since this description is closer to the experimental setup we propose. Since our approach uses very different methods than the one employed in  \cite{Damour:1995kt}, a general comparison of the results is rather involved and beyond the scope of this article.

We will first linearize the theory around Minkowski spacetime in section \ref{sec:weakST} before we solve the resulting linearized field equations sourced by an energy-momentum tensor of an ultrarelativistic particle in section \ref{sec:URsource}. Finally we derive the effect on a test mass that acts as a sensor, such as an opto-mechanical detection device, in section \ref{sec:sensor} and discuss the outcome of such an experiment in section \ref{sec:test}, before we conclude in section \ref{sec:conc}.

Throughout the article, we assume that the Minkowski metric takes the form $\eta=\mathrm{diag}(-1,1,1,1)$. Indices $J,K, ...$ run over $1,2$, indices $\mu,\nu,\sigma, ...$ run over $0,1,2,3$.

\section{The weak field limit of scalar tensor gravity}\label{sec:weakST}

Scalar tensor theories aim to describe the gravitational field in terms of a metric $g$ with components $g_{\mu\nu}$ and an additional scalar field $\Psi$. In the literature, there are various different realisations of scalar tensor theories. Many of them can be captured by the following action (sometimes called Bergman-Wagoner theories \cite{Bergmann:1968ve,Wagoner_1970}), which is a generalisation of the Brans-Dicke theory, with a free coupling function $\omega(\Psi)$ and a free scalar field potential $V(\Psi)$ \cite{Esposito-Farese:2000pbo,Flanagan:2004bz,Hohmann:2013rba},
\begin{widetext}
\begin{equation} \label{eqn:action}
S[g_{\mu\nu},\Psi,\chi_m] = \frac{1}{2\kappa^2}\int_{M}d^4x\sqrt{-\det g}\left(\Psi R - \frac{\omega(\Psi)}{\Psi}\partial_{\rho}\Psi\partial^{\rho}\Psi - 2\kappa^2V(\Psi)\right) + S_m[g_{\mu\nu},\chi_m]\,.
\end{equation}
\end{widetext}
where $\kappa$ is the bare gravitational constant of the theory, $R$ is the Ricci scalar of the Levi-Civita connection of the metric, $\det g$ is the determinant of the metric, and $S_m$ the matter field action for various matter fields summarized as $\chi_m$.
The choice of the coupling function and the potential fixes different theories. To avoid pathologies such as ghost modes, one restricts to theories which satisfy $2 \omega(\Psi)+3 > 0$ and which have a potential bounded from below  $V(\Psi)>C \in \mathbb{R}$.

In addition to various scalar tensor theories this action also captures the scalar tensor representation of the famous $f(R)$ theories \cite{Teyssandier:1983zz,Faulkner:2006ub}, which are important in the context of cosmology and inflation \cite{CANTATA:2021ktz}. More explicitely, the action
\begin{align}
    S[g_{\mu\nu},\chi_m] = \int d^4x  \frac{1}{2\kappa^{2}} \sqrt{-\det g} f(R) + S_M[g_{\mu\nu},\chi_m] 
\end{align}
is equivalent to the scalar tensor action with non-minimally coupled scalar field $\Psi$ and minimal coupling to the matter sector,
\begin{equation}
\begin{aligned}
    S[g_{\mu\nu},\Psi,\chi_m] 
    &= \int d^4x  \sqrt{-\det g} \left(\frac{1}{2\kappa^{2}}  \Psi  R -  V(\Psi) \right) \\
    & \quad + S_M[g_{\mu\nu},\chi_m]\,,
\end{aligned}
\end{equation}
where $f'(R)=\Psi$ and $V(\Psi) = \frac{1}{2\kappa^2} (R(\Psi)\Psi - f(R(\Psi)))$. For example, the famous Starobinsky inflation model $f(R)=R+\lambda R^2$ is included in the classes of theories we are considering by setting $\omega=0$ and realizing that $f'(R) = 2\lambda R = \Psi$ and thus $V(\Psi) = \frac{1}{2\kappa^2} \frac{(\Psi-1)^2}{2\lambda}$. In this way, the test of scalar tensor theories with ultrarelativistic particles, which we propose and discuss in detail the following, can also be used to test $f(R)$ theories, and thus Starobinsky inflation, in a completely complementary local environment, compared to the usual tests on cosmic scales.

In order to analyse the ultrarelativistic limit of the theories defined by \eqref{eqn:action}, which is a regime that is not much investigated compared to their post-Newtonian limit or their impact on cosmology, we quickly recall the field equations. Variation of the action with respect to $g_{\mu\nu}$ and $\Psi$ (and employing a partial decoupling by eliminating the Ricci scalar from the field equations with help of the trace of the metric field equation) leads to the following form of the field equations, see \cite{Hohmann:2013rba} for more details. For the metric one gets
\begin{eqnarray}
R_{\mu\nu} &=& \frac{1}{\Psi}\Bigg[\kappa^2\left(T_{\mu\nu} - \frac{\omega + 1}{2\omega + 3}g_{\mu\nu}T\right) + \nabla_{\mu}\partial_{\nu}\Psi \nonumber \\
&& \quad \quad + \frac{\omega}{\Psi}\partial_{\mu}\Psi\partial_{\nu}\Psi - \frac{g_{\mu\nu}}{4\omega + 6}\frac{d\omega}{d\Psi}\partial_{\rho}\Psi\partial^{\rho}\Psi\Bigg] \label{eqn:metric_main} \\
&&+ \frac{\kappa^2}{\Psi}{g_{\mu\nu}}\frac{2\omega + 1}{2\omega + 3}V + g_{\mu\nu}\frac{\kappa^2}{2\omega + 3}\frac{dV}{d\Psi}\,,\nonumber
\end{eqnarray}
and for the the scalar field
\begin{eqnarray}
    \largesquare\Psi &=& \frac{1}{2\omega + 3}\left[\kappa^2T - \frac{d\omega}{d\Psi}\partial_{\rho}\Psi\partial^{\rho}\Psi + 2\kappa^2\left(\Psi\frac{dV}{d\Psi} - 2V\right)\right]\,,\label{eqn:scalar_main}
\end{eqnarray}
where $\omega=\omega(\Psi)$, $V=V(\Psi)$, $\largesquare = g^{\mu\nu}\nabla_\mu\nabla_\nu$ is the d'Alembert operator on curved spacetimes, $R_{\mu\nu}$ the Ricci-Tensor of the Levi-Civita connection, $T_{\mu\nu}$ is the energy momentum tensor and $T$ its trace. The first step towards deriving the gravitational field of moving point sources is to linearize the field equations. We employ the ansatz
\begin{align}
    g_{\mu\nu} &= \eta_{\mu\nu} + h_{\mu\nu}\,,\label{eq:linmet}\\
    \Psi &= \psi_0 + \psi_1\,,\label{eq:linscal}
\end{align}
where we assume that the perturbations $h_{\mu\nu}$ and $\psi_1$ as well as the components $T_{\mu\nu}$ of the energy-momentum tensor are of the same small order.
We now expand the field equations to first order in all of the just mentioned quantities. 

To solve the field equations consistently order by order for the perturbative ansatz \eqref{eq:linmet} and \eqref{eq:linscal}, the background scalar field $\psi_0$ must satisfy two equations: the zeroth order wave equation \eqref{eqn:scalar_main} and a constraint from the zeroth order of the metric field equation \eqref{eqn:metric_main} whose left hand side vanishes. This implies that in particular the trace of the right hand side of \eqref{eqn:metric_main} must vanish at zeroth order. Assuming that on Minkowski spacetime there is no gravitational effect at all, also not from the additional scalar field, we set $\partial_\mu \psi_0 = 0$ everywhere and $\psi_0\neq0$ which solves the field equations when the potential is chosen such that $V'(\psi_0)=0$ and $V(\psi_0)=0$. These choices are satisfied for many models discussed in the literature.

To further analyse the first order field equations, we define the background values $\omega_0 = \omega(\psi_0)$, $V''_0=V''(\psi_0)$ and the trace-reversed scalar field shifted metric perturbation $\bar{h}_{\mu \nu }=h_{\mu \nu } - \eta_{\mu \nu }(\psi_{1}{}/\psi_{0}{}+h/2)$, where $h=\eta^{\mu\nu}h_{\mu\nu}$, and the effective scalar field mass $m_{\psi_1}=\sqrt{2\kappa^2 V''_0 \psi_0/(3 + 2 \omega_0)}$. Moreover, we employ the scalar-tensor theory adopted harmonic gauge condition $\partial_\mu h^\mu{}_\nu = \tfrac{1}{2}\partial_\nu h^\alpha{}_\alpha + \frac{1}{\psi_0}\partial_\nu \psi_1$, see also \cite{Nutku1969}, which implies for the trace reverse metric $\partial_\nu \bar h^\nu{}_\mu = 0$. Observe that, at first order, indices are raised and lowered with the Minkowski metric. 

The resulting linearized field equations can now be found to be
\begin{align}\label{eq:metricwaveeq}
    \eta^{\alpha\beta}\partial_{\alpha }\partial_{\beta}\bar{h}_{\mu \nu } 
    &= -\frac{2 \kappa^2}{\psi_{0}{}} T_{\mu \nu }   \\
    \label{eq:scalarwaveeq}
    \eta^{\alpha\beta}\partial_{\alpha }\partial_{\beta}\psi_{1} - m_{\psi_1}^2 \psi_{1}{}  &= \frac{\kappa^2 T}{3+2\omega_0}\,.
\end{align}
In this form the theory either contains three free parameters $\psi_0$, $\omega_0$ and $m_\psi$ or one fixes the theory in advance by specifying the functions $V(\psi)$ and $\omega(\psi)$, then only the background value $\psi_0$ has to be determined as free parameter. We will see that the solutions to the equations have a smooth $m_{\psi_1} \to 0$ limit.

Note that the linearized Einstein equations are recovered for $\psi_{0}{}=1$ and $\bigl(3 + 2 \omega_0\bigr)/(1 + \omega_0) = 2$, that is, $\omega_0\rightarrow \infty$. Below, we will use this limit to recover the gravitational field of moving point-like sources in GR.

\section{The gravitational field of a moving point-like source}\label{sec:URsource}
To solve the field equations for the linearized gravitational field \eqref{eq:metricwaveeq} and \eqref{eq:scalarwaveeq} of ultrarelativistic particles, we need to specify the energy momentum tensor. 

In this article, we consider a single 
point particle. Our results can then be used to model sources that are collections of free point particles such as the proton bunches in the beamline at the LHC. We consider that the particle has a mass $M$ and moves with constant velocity $v$ in the $z$-direction, that is, its lab-frame velocity vector is given as $(v^0,v^1,v^2,v^3)=(c,0,0,v)$.
 
Then, the energy momentum tensor in terms of the particle's mass and velocity and local laboratory coordinates $(ct,x,y,z)$ is given by
\begin{equation}\label{eq:EMparticles2}
	T_{\mu\nu} = \frac{u(\vec{r},t)}{c^2} v_\mu v_\nu\,,
\end{equation}
where $u(\vec{r},t)=E\delta(z-vt)\delta(x)\delta(y)$, $E=M\gamma c^2$ and $\gamma = (1-v^2/c^2)^{-1/2}$. We recall that this energy-momentum tensor is of first order, and thus its trace is, to leading order,
\begin{equation}
	T = -\frac{u(\vec{r},t)}{\gamma^2}\,.
\end{equation}
The solutions of equations \eqref{eq:metricwaveeq} and \eqref{eq:scalarwaveeq} are then derived in a straight forward fashion, see appendix~\ref{app:fieldEQ} for details, for example, by the method of retarded potentials or by applying a Lorentz boost to the corresponding solutions for a source at rest (which can be found e.g. in \cite{Hohmann:2013rba}). Using the abbreviation $\rho^2 = x^2 + y^2$, the solution for the scalar field is 
\begin{equation}\label{eq:psi1}
    \psi_{1} = \frac{\kappa^2}{4\pi(3+2\omega_0)} \frac{M c^2}{\sqrt{\gamma^2(z-vt)^2+\rho^2}} e^{-m_{\psi_1}\sqrt{\gamma^2(z-vt)^2+\rho^2}}\,,
\end{equation}
and the for the metric perturbation, we obtain 
\begin{equation}\label{eq:metrich}
\begin{aligned}
      h_{\mu \nu } 
      &= \bar  h_{\mu\nu} - \eta_{\mu\nu} \left(\frac{\psi_1}{\psi_0} + \frac{\bar h}{2}\right)\\
      &= \frac{\gamma^2}{c^2} v_\mu v_\nu \mathcal{K}  + \tfrac{\eta_{\mu \nu }}{2} \mathcal{H} \,,
\end{aligned}
\end{equation}
where 
\begin{align}
	\mathcal{K} &= \frac{2\kappa^2}{4\pi\psi_{0}}\frac{M c^2}{\sqrt{\gamma^2(z-vt)^2+\rho^2}} \label{eq:defK}\\
	\mathcal{H}  &=
 \left( 1 - \tfrac{e^{-m_{\psi_1}\sqrt{\gamma^2(z-vt)^2+\rho^2}}}{3+2\omega_0}\right) \mathcal{K}\,.\label{eq:defH}
\end{align}
Note that the solutions have a smooth $m_{\psi_1}\to 0$ limit, hence can be used for massless and massive scalar fields at the same time, by choosing the mass parameter accordingly.
In the limit $m_{\psi_1}\to 0$, the expression of the metric perturbation depends on the scalar field theory only via $\psi_0$ and $\omega_0$. Since $\psi_0$ can be absorbed into the gravitational constant (see Sec. \ref{sec:test}), the only remaining parameter is $\omega_0$.

GR corresponds to the case $\mathcal{H}\to \mathcal{K}$ (or $\omega_0 \to 
\infty$). Moreover the derivatives of these function satisfy the convenient identity
\begin{equation}\label{eq:dtKtod3K}
	\partial_t \mathcal{K}  = - v\partial_3 \mathcal{K} \quad \text{and}\quad \partial_t \mathcal{H}  = - v\partial_3 \mathcal{H}\,,
\end{equation}
with, for $v\neq0$, the important properties
\begin{align}
    \lim_{t\to\pm\infty} \partial_t \mathcal{K} &= \lim_{t\to\pm\infty} \partial_K \mathcal{K} = \lim_{t\to\pm\infty} \mathcal{K} = 0\,,\label{eq:limK}\\
    \lim_{t\to\pm\infty} \partial_t \mathcal{H} &= \lim_{t\to\pm\infty} \partial_K \mathcal{H} =  \lim_{t\to\pm\infty} \mathcal{H} = 0\,.\label{eq:limH}
\end{align}

\section{Effect on an acceleration sensor}\label{sec:sensor}
Having found the gravitational field of the point particle source 
in scalar tensor theory, we now investigate how test particles would be affected.

To do so, we calculate the spatial coordinate acceleration $a_T$ of test particles from the geodesic equation in Section \ref{ssec:spatacc}. We focus on the leading-order terms in the test particles’ velocity $v_T$, as we assume that the test particles are initially at rest. Furthermore, we consider only those terms that are not proportional to the $t$- and $z$-derivatives of $\mathcal{K}$ and $\mathcal{H}$, since only these terms contribute to the momentum transfer to the test particle at rest. This momentum transfer will be derived by integrating the acceleration from $t = -\infty$ to $t = \infty$ in Section \ref{ssec:mtransf}. In Section \ref{ssec:geoddevi}, we discuss why the transverse coordinate acceleration is the correct quantity to use for deriving the momentum transfer, as it can be obtained from the geodesic deviation equation, even though itself is not a tensorial quantity.

\subsection{The spatial acceleration}\label{ssec:spatacc}
The spatial coordinate acceleration $a_T$ of a test mass probing the gravitational field of the particle beam is derived from the geodesic equation of the metric \eqref{eq:linmet}
($i,j,k, ... = 1,2,3$),
\begin{align}\label{eq:spatacc1}
    \frac{d^2 x_T^i}{dt^2} 
    & = a_T^i  = - c^2\Gamma^i{}_{00} + \mathcal{O}\left(\tfrac{v_T}{c}\right) \nonumber \\
    &= \frac{c}{2}\eta^{ij} \big(c\ \partial_j h_{00} - 2 \partial_t h_{j0} \big) + \mathcal{O}\left(\tfrac{v_T}{c}\right)\,,
\end{align}
where, in the last line, we displayed the leading order terms in the velocities of the test particle $v_T$ expressed in terms of the metric components to first order. Our focus lies on the leading order effect, i.e.\ the terms not involving any factor of $v_T$, as our test particles will be slow. The use of relativistic test particles ($v_T\gg0$) is much more challenging and might can be investigated in the future.
\begin{figure*}
    \includegraphics[width=4.6cm,angle=0]{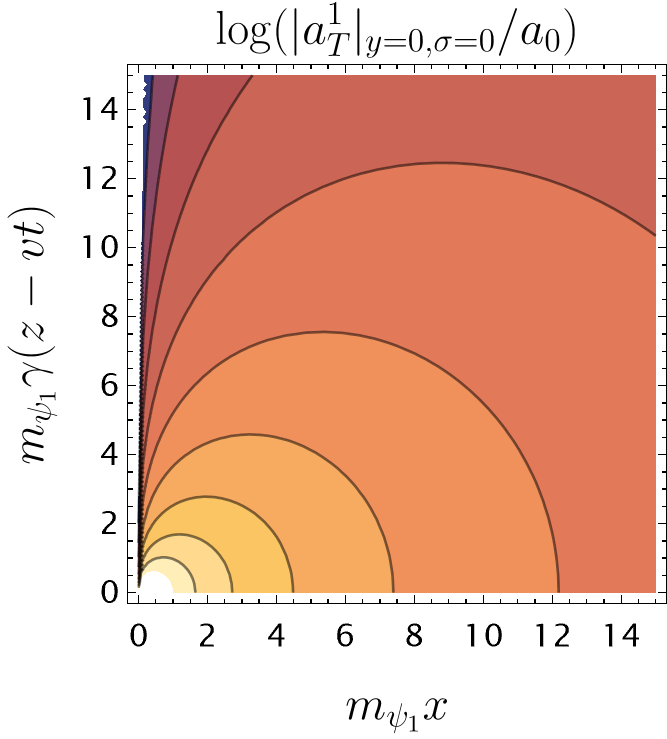}
    \includegraphics[width=5.5cm,angle=0]{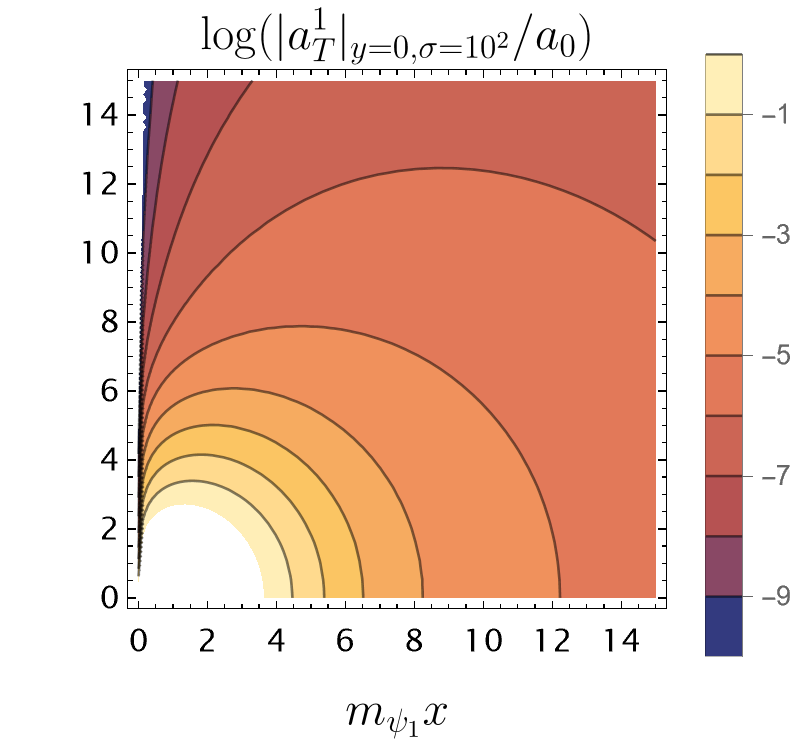}
    \caption{These plots show the logarithm of the absolute value of the acceleration in $x$-direction $a_T^1$ of a test particle at rest at $y=0$ as a function of $m_{\psi_1}x$ and $m_{\psi_1}\gamma(z-vt)$ in units of $a_0=\kappa^2Mc^4(\gamma^2-1/2)m_{\psi_1}^2/(4\pi\psi_0)$ for two different values of $\sigma=(2(\gamma^2-1/2)(3+2\omega_0))^{-1}$.
    Left: $\sigma=0$, equivalent to the GR case times a global factor $\psi_0^{-1}$. Right: $\sigma=10^2$ on the right hand side. 
    The two plots agree for large distances, while the modification due to the scalar field dominate at small distances.
    \label{fig:acc-plot}}
\end{figure*}

To derive the magnitude of the leading order spatial acceleration, we thus need the derivatives, $\partial_j h_{00}$ and $\partial_t h_{j0}$ which we display in appendix \ref{app:subleading}. It turns out that for the momentum transfer to the test particle, which we derive in section \ref{ssec:mtransf}, only the acceleration $a_T^K$ transverse to the beam-line is relevant, since $\partial_t h_{j0}$ and $\partial_3 h_{00}$ are proportional to $\partial_t \mathcal{K}$ or $\partial_t \mathcal{H}$, and thus vanish when integrated from $t=-\infty$ to $t=\infty$. Thus, the $x$- and $y$-derivatives of $h_{00}$ determine the leading order acceleration that affects a non-relativistic test particle at rest in the lab frame (i.e. with initial velocity vector $(c,0,0,0)$ with respect to the coordinates we use for our calculations). The acceleration is transverse to the beam-line and takes the form
\begin{widetext}
\begin{equation}
    \begin{aligned}\label{eq:atrans1}
        a_T^{K} &= \frac{c^2}{2}\partial^K h_{00} 
        = -\frac{\kappa^2M c^4}{4\pi\psi_{0}} \frac{
        x^K
        }{(\gamma^2(z-vt)^2+\rho^2)^{3/2}} \left( \gamma^2  - \frac{1}{2}  + \frac{1 + m_{\psi_1}\sqrt{\gamma^2(z-vt)^2+\rho^2}}{2(3+2\omega_0)}  e^{-m_{\psi_1}\sqrt{\gamma^2(z-vt)^2+\rho^2}}\right)   \,.
    \end{aligned}
\end{equation}
\end{widetext}
In Fig. \ref{fig:acc-plot}, we present plots of the absolute value of the acceleration for different parameter values. The GR case can easily be obtained by performing the necessary limits $\psi_0=1$ and $\omega_0 \to \infty$, as noted above, leading to
\begin{eqnarray}\label{eq:accGR}
	  a^K_{T,\mathrm{GR}} &=& -\frac{\kappa^2 M c^4}{4\pi} \frac{x^K
   }{(\gamma^2(z-vt)^2+\rho^2)^{3/2}}\left(\gamma^2-\frac{1}{2}\right)  \,.
\end{eqnarray}
Note that the coordinate dependence of the GR acceleration agrees with that of \eqref{eq:atrans1} in the limits, $m_{\psi_1}\to 0$ and $m_{\psi_1}\to \infty$, while only in the limit $m_{\psi_1}\to \infty$, also the velocity dependence coincides.

Besides the leading order transversal acceleration, there exist a sub-leading order transversal component of the acceleration of test particles, which are caused by the gravitomagnetic part of the gravitational field of the ultra-relativistic particle. We discuss their precise expression and magnitude in appendix~\ref{app:trans}. Moreover, there exist longitudinal components of the acceleration of test particles, which we discuss in appendix~\ref{app:longi}. The leading order terms are all proportional to the $t$- and $z$-derivatives of $\mathcal{H}$, $\mathcal{K}$, which is why they do not contribute in the momentum transfer to the test particle. In addition there are sub-leading terms, partly also sourced by gravitomagnetism.

\subsection{The geodesic deviation}\label{ssec:geoddevi}
Note that acceleration as expressed in \eqref{eq:spatacc1} is a coordinate-dependent non-tensorial quantity, and therefore, does usually not serve as a physical observable in the context of GR. Instead, one has to study relative quantities, for example, the relative acceleration of two freely falling neighboring test particles, described by two geodesics $\gamma$ with parallel tangents $\dot\gamma$ displaced by a deviation vector $s$. The general expression of this differential acceleration, the geodesic deviation, is
\begin{equation}
    \ddot s^\mu = {R^\mu}_{\nu\rho\sigma} \dot\gamma^\nu\dot\gamma^\rho s^\sigma\,.
\end{equation}
where ${R^\mu}_{\nu\rho\sigma}$ is the Riemann curvature tensor which takes the following form in first order in the metric perturbation
\begin{equation}\label{eq:lincurv}
	\begin{aligned}
    {R^{\mu}}_{\beta\gamma\delta} &= \frac{\eta^{\mu\alpha}}{2}\left(\partial_\beta\partial_\gamma h_{\delta\alpha} - \partial_\beta\partial_\delta h_{\gamma\alpha} - \partial_\gamma\partial_\alpha h_{\beta\delta} + \partial_\delta\partial_\alpha h_{\beta\gamma}\right) \,.
    \end{aligned}
\end{equation}
We give expressions for the curvature tensor in appendix~\ref{app:Curv}. 

For the transverse differential acceleration of geodesics that differ initially by $s=(0,s^1,s^2,0)$, we obtain, to leading (zeroth) order in the their initial velocity $\dot\gamma^j$, 
\begin{align}\label{eq:ddotsJ}
    \ddot s^J 
    = c^2 {R^J}_{00K}s^K
    &=  s^K \left(\partial_K a_T^J + \delta^J_K\frac{1}{4}\partial_t\partial_t \mathcal{H}\right) 
\end{align}
where we used the first equality in \eqref{eq:atrans1} and \eqref{eq:metrich}.

In the large $v$ limit, the second term in equation \eqref{eq:ddotsJ} gives a significant contribution to the acceleration. However, we find that it does not contribute to the time integrated differential acceleration to leading order as it is a second time derivative and the first time derivative of $\mathcal{H}$ vanishes for $t=\pm\infty$ and finite $z$. We conclude that the second term in $\ddot s^J$ is only relevant for detectors that can resolve the temporal dependence of the acceleration, which is characterized by the time scale $t_\mathrm{pulse}\sim \rho/(\gamma v)$, given by the decay of the gravitational field with powers of the inverse of $\sqrt{\rho^2-\gamma^2(vt)^2}$. Hence, focusing only on detectors that are unable to resolve $t_\mathrm{pulse}$ in the following, we can associate an observable quantity with the transverse acceleration. 

Similarly, the integrated longitudinal differential acceleration of geodesics vanishes (the derivation can be found in appendix \ref{app:geoddevi}), which implies that also the longitudinal acceleration can be neglected for detectors unable to resolve $t_\mathrm{pulse}$.

\subsection{The momentum transfer}\label{ssec:mtransf}
As discussed above, the relevant leading order  term for the differential acceleration is the gradient of the transverse acceleration $a^K$, the first term in \eqref{eq:ddotsJ}. Integrating it along the transverse directions, i.e. to the distance of the sensor from the beam-line, shows that the important quantity for the motion of the sensor is the transverse test particle acceleration $a_T^K$ itself.
 
\begin{figure}
    \includegraphics[width=6cm,angle=0]{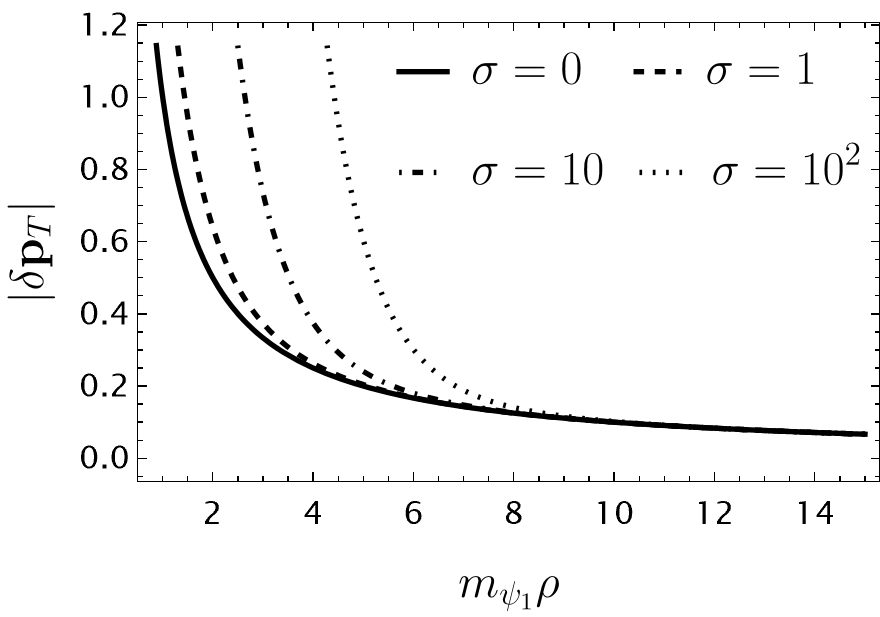}
    \caption{This plot shows the absolute value of the momentum transfer $\delta\vec{p}_T$ of a test particle at rest as a function of $m_{\psi_1}\rho$ in units of $\kappa^2M m c^4(\gamma^2-1/2)m_{\psi_1}/(2\pi\psi_0\gamma v)$ for different values of $\sigma=(2(\gamma^2-1/2)(3+2\omega_0))^{-1}$. For $\sigma=0$ and for large distances, the $1/\rho$-scaling of GR is recovered. For $\sigma>0$ and small distances, the modification due to the scalar field dominates. \label{fig:mtrans-plot}}
\end{figure}

To discuss this leading order effect, we note that a quickly moving source will only be slightly deflected from its path by the back-action from an acceleration sensor and the acceleration is concentrated in a pulse of duration $t_\mathrm{pulse}$ around $t=z/v$. In particular, for the ultrarelativistic regime and short distances to the beam line, this time-scale is quite small, for example, $~100\,$fs for $\rho\sim 1\,$cm and $1 - v/c \sim 10^{-6}$ corresponding to injection speed at the LHC.
Thus, we can approximate the transverse momentum transfer by integrating the transverse acceleration over time from $-\infty$ to $\infty$ at the initial position $\vec{x}$ of the test particle, and assuming that the trajectory of the test particle sensor is only disturbed slightly as (in Minkowski coordinates) $\gamma^\mu = (c t, x,y,z) + \Delta \gamma^\mu$, where $\Delta \gamma^\mu$ is of first order in the metric perturbation. We obtain

\begin{equation}
    \begin{aligned}\label{eq:deltapscalartensor}
         \delta p^K_T &\sim  m \int_{-\infty}^\infty dt\, a^K_T(t,\vec{x}) \\
         & = -\frac{ \kappa^2 M m c^4}{4\pi\psi_0 } \frac{x^K}{\gamma v} \Bigg( \left(\gamma^2 - \frac{1}{2}\right)\frac{2}{\rho^2} \\
         & \quad +\frac{1}{(3+2\omega_0)}\frac{m_{\psi_1}}{\rho}    K_1(m_{\psi_1}\rho) \Bigg)\,.
    \end{aligned}
\end{equation}
Here, $m$ is the test mass of the sensor and $K_1$ is a modified Bessel functions of the second kind. A plot of the momentum transfer for different values of the parameters is shown in Fig. \ref{fig:mtrans-plot}.  Again, we recover the GR case in the limit $\psi_0=1$ and $\omega_0 \to \infty$, leading to
\begin{equation}
	\begin{aligned}\label{eq:deltapGR}
	\delta p^K_{T,\mathrm{GR}} &\sim -\frac{\kappa^2 M m c^4 }{2\pi}  \frac{x^K}{\gamma v \rho^2}\left(\gamma^2-\frac{1}{2}\right)   \,.
	\end{aligned}
\end{equation}
Note that the differences between the scalar-tensor case and GR are a global factor $\psi_0^{-1}$ (which is un-observable as it can be absorbed in the measured gravitational constant, see Sec. \ref{sec:test}) and an additional term contributed by the scalar field proportional to $K_1(m_{\psi_1}\rho)$, which is suppressed in the ultrarelativistic regime. In the next section, we will discuss explicitly how one could use these differences to test scalar-tensor theories with the gravitational field of ultrarelativistic particles. 

\section{Test Scenario and Discussion}\label{sec:test}
Having identified the momentum transfer to the test sensor from the ultra-relativstic particle via its gravitational field, we now discuss how this effect can be compared with the expectations based on general relativity. First we discuss the general procedure before we evaluate our findings for a particle beam at the LHC.

\subsection{The general test scenario}
In an experimental test of GR, the expressions in equations \eqref{eq:accGR} and \eqref{eq:deltapGR} represent the expectations with which one would compare the experimental results. However, $\kappa$ is a bare parameter that is not known a priori. To obtain the null-hypothesis of a potential test, $\kappa$ has to be substituted by a $\tilde{\kappa}$ for which an appropriate measurement prescription is given. 

Conventionally, the gravitational constant is measured by highly optimized experiments in the non-relativistic regime, for example, by a torsion balance. Hence, we assume that the measurement of $\tilde{\kappa}$ is performed with slowly moving source and test masses at a distance $r_0 = |\vec{x}_0|$ (Minkowski coordinate distance) through an acceleration measurement, that is, the value of the $\tilde{\kappa}_{r_0}$ is inferred from the Newtonian limit expectation:
\begin{equation}
    |\vec{a}_\mathrm{measured}|_{r_0} = \frac{\tilde{\kappa}_{r_0}^2 Mc^4}{8\pi r_0^2}\,.
\end{equation} 
The bare gravitational constant $\kappa$ that has to be used for predictions of scalar-tensor theory can be derived by comparing the above expression with \footnote{Note that the right hand side of equation \eqref{eq:acc-comp} shows the Yukawa form of the acceleration in scalar-tensor theories. We can read off the commonly used parameters of Yukawa-type forces as $\alpha = 1/(3+2\omega_0)$ and $\lambda = 1/m_{\psi_1}$. For these parameters, stringent bounds exist \cite{Murata_2015,Tan2020:imp}.} 
\begin{eqnarray}\label{eq:acc-comp}
    |\vec{a}_{T,\mathrm{slow}}|_{r_0} &=& \left|\frac{c^2}{2}\vec{\partial} h_{00}\right|_{v=0,r_0} \\
    \nonumber &=&  \frac{\kappa^2M c^4}{8\pi\psi_{0}r_0^2} \left( 1 + \frac{1 + m_{\psi_1} r_0}{3+2\omega_0}  e^{-m_{\psi_1}r_0}\right)  
\end{eqnarray}
leading to 
\begin{equation}\label{eq:tildekappaofr0}
     \kappa(\tilde{\kappa}_{r_0})^2 = \psi_{0}\tilde{\kappa}_{r_0}^2 \,\left( 1 + \frac{1 + m_{\psi_1}r_0}{3+2\omega_0}e^{-m_{\psi_1}r_0}\right)^{-1}\,.
\end{equation}

\begin{figure}
    \vspace{10mm}
    \includegraphics[width=4cm,angle=0]{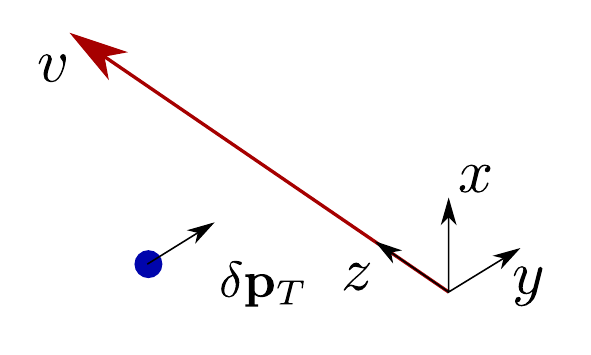}
    \vspace*{-3mm}
    \caption{Sketch of the test-setup: measurement of the momentum transfer on a test particle in the gravitational field of a relativistic particle which moves with velocity $v$ in the $z$-direction.
    \label{fig:setup}}
\end{figure}
Therefore, the difference between the momentum transfer and the expected momentum transfer based on the GR prediction with $\tilde{\kappa}_{r_0}$ becomes
\begin{widetext}
    \begin{equation}
    \begin{aligned}\label{eq:Deltadeltap}
        \Delta \delta p^K_T 
        = \delta p^K_T - \delta p^K_{T,\mathrm{GR}}(\tilde{\kappa}_{r_0}) 
        =  -\frac{\kappa(\tilde{\kappa}_{r_0})^2 M m c^4}{4\pi\psi_0 (3+2\omega_0)} \left( -\left(\gamma^2 - \frac{1}{2}\right)\frac{2}{\rho^2}(1 + m_{\psi_1}r_0)e^{-m_{\psi_1}r_0} +\frac{m_{\psi_1}}{\rho}    K_1(m_{\psi_1}\rho) \right)\frac{x^K}{\gamma v} \,.
    \end{aligned}
\end{equation}
\end{widetext}
Since the expressions for the momentum transfer in equations \eqref{eq:deltapscalartensor}, \eqref{eq:deltapGR} and \eqref{eq:Deltadeltap} are proportional to coordinate positions in Minkowski spacetime, the difference of the absolute values of the momentum transfer is just the absolute value of the difference of the vectorial momentum transfer, that is, $\Delta \delta p_T = |\Delta \delta p^j_T|$. Therefore, we can write the relative momentum transfer normalized by $\delta p_{T,\mathrm{GR}}(\tilde{\kappa}_{r_0})$ as
\begin{eqnarray}\label{eq:relmomentumtransmassive_main}
    \frac{\Delta \delta p_T}{\delta p_{T,\mathrm{GR}}(\tilde{\kappa}_{r_0})} & = &\frac{\kappa(\tilde{\kappa}_{r_0})^2}{\psi_0\tilde{\kappa}_{r_0}^2} \Bigg| \frac{1 + m_{\psi_1}r_0}{3+2\omega_0}e^{-m_{\psi_1}r_0} \\
    \nonumber && \quad - \frac{m_{\psi_1}\rho}{2(3+2\omega_0)}  \left(\gamma^2 - \frac{1}{2}\right)^{-1}  K_1(m_{\psi_1}\rho) \Bigg|
\end{eqnarray}
Note again that the additional length scale $r_0$ arises due to the estimation of the gravitational constant in an auxiliary experiment with non-relativistic source and test masses at fixed distance $r_0$. 

\begin{figure}
    \includegraphics[width=6cm,angle=0]{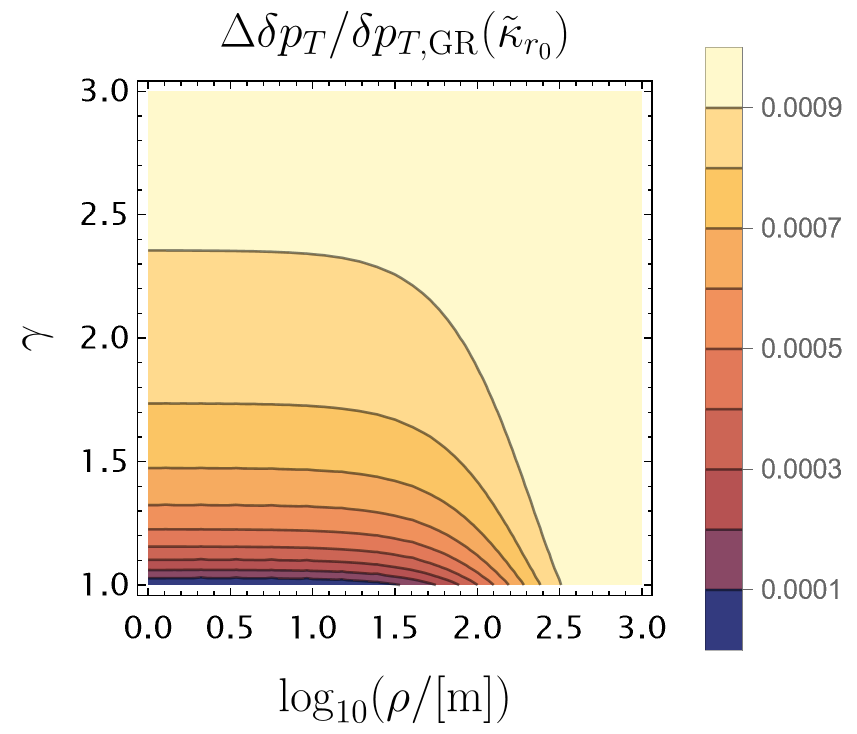}
    \caption{This plot shows the relative momentum transfer difference due to a moving source defined as the difference between the values of the transverse momentum transfer to a test particle derived from scalar-tensor theory and GR normalized by the latter. The values are given for different values of the Lorentz factor $\gamma$ and the distance between test particle and beam line $\rho$. For the plot, the scalar field parameters were chosen as $1/(3+2\omega_0)=10^{-3}$ and $1/m_{\psi_1}=10^2\,$m which are below the bounds imposed by experimental and observational tests of Yukawa type modifications of gravity (compare with Fig. 8 of \cite{Murata_2015} taking into account that $\alpha=1/(3+2\omega_0)$ and $\lambda= 1/m_{\psi_1}$). Furthermore, for the distance at which the gravitational constant is fixed by auxiliary experiments, we have used $r_0=0.1\,$m, while the plots look equivalent for other values of $r_0\lesssim 1\,$m. The plot clearly shows the advantage of tests with relativistic sources if $\rho$ and $r_0$ are small in comparison to $1/m_{\psi_1}$.
    \label{fig:rel-plot}}
\end{figure}

We find that the right hand side of \eqref{eq:relmomentumtransmassive_main} contains two terms: the first decays exponentially with $m_{\psi_1}r_0$ and the second decays exponentially with $m_{\psi_1}\rho$. Hence, the first term is negligible when the gravitational constant is fixed at distances $r_0\gg 1/m_{\psi_1}$, and the second term is negligible for $\rho\gg 1/m_{\psi_1}$. This is just the usual result of an exponential scaling with distance that is common to every Yukawa-type potential. Thus, in general, some of the steps of the test have to be performed at distances close to $1/m_{\psi_1}$, either the fixing of the gravitational constant or the measurement of the gravitational effect of the moving source mass on a sensor at rest, or both. To investigate quantitatively the behaviour of the relative momentum transfer difference in \eqref{eq:relmomentumtransmassive_main} as function of the source particle velocity and distance $\rho$, we plot its behaviour in Fig. \ref{fig:rel-plot} for parameters of the scalar-tensor theory that are not excluded by present bounds \cite{Murata_2015}. We clearly see the strong impact of the velocity of the gravitational source for $\rho\ll 1/m_{\psi_1}$, the larger the velocity the larger is the relative momentum transfer difference. Hence, the velocity of the source introduces a new parameter that can be used for novel tests. 

The analytic expression can be investigated for two regimes. First, in the case that $m_{\psi_1}r_0\gg 1$, that is, when the gravitational constant is fixed by measurement at large distances, we obtain
\begin{equation}
    \frac{\Delta \delta p_T}{\delta p_{T,\mathrm{GR}}(\tilde{\kappa}_{r_0})} \xrightarrow{m_{\psi_1}r_0\to\infty} \frac{m_{\psi_1}\rho}{2(3+2\omega_0)}  \left(\gamma^2 - \frac{1}{2}\right)^{-1}  K_1(m_{\psi_1}\rho) 
\end{equation}
This decreases monotonically with $\gamma$ and there is no advantage of performing an experiment with moving source masses besides performing an independent test. Second, in the case $m_{\psi_1}\rho\gg 1$, i.e. for large distances between the moving source mass and the sensor, we find
\begin{equation}\label{eq:relmomentumtransmassivelim}
    \frac{\Delta \delta p_T}{\delta p_{T,\mathrm{GR}}(\tilde{\kappa}_{r_0})} \xrightarrow{m_{\psi_1}\rho\to \infty} \frac{\kappa(\tilde{\kappa}_{r_0})^2}{\psi_0\tilde{\kappa}_{r_0}^2} \frac{1 + m_{\psi_1}r_0}{3+2\omega_0}e^{-m_{\psi_1}r_0} \,.
\end{equation}
This expression does not contain any dependence on the velocity of the source mass, and therefore, there is again no advantage of employing moving sources for the detection of a scalar field modification of GR in this regime. We see again that both $r_0$ and $\rho$ have to be smaller or of the same magnitude as $1/m_{\psi_1}$ to gain an advantage from moving (relativistic) source masses. 

The most extreme situation where these conditions are satisfied corresponds to the massless scalar field limit $m_{\psi_1}\to 0$. Then, we find 
\begin{equation}
    \begin{aligned}
	\frac{\Delta \delta p_T}{\delta p_{T,\mathrm{GR}}(\tilde{\kappa}_{r_0})} &\approx \frac{1}{2(2+\omega_0)} \, \frac{\gamma^2 - 1}{\gamma^2-\frac{1}{2}} = \frac{1}{2+\omega_0} \, \frac{v^2}{v^2+c^2}  \,.
    \end{aligned}
\end{equation}
We see that there is a velocity-dependent momentum transfer difference which depends on $\omega_0$ as the only other parameter and saturates at a value $1/2(2+\omega_0)$ in the ultrarelativistic regime 
\footnote{ Interestingly, this post-Minkowskian result can be interpreted in terms of the parametrised post-Newtonian (PPN) parameter $\gamma^{\textrm{PPN}}$, which is related to $\omega_0$ as $\omega_0=(2 \gamma^{\textrm{PPN}}-1)/(1- \gamma^{\textrm{PPN}})$ (see \cite{Hohmann:2013rba} where $\gamma^{\textrm{PPN}}$ is just denoted as $\gamma$). The relative momentum transfer difference for the case of a massless scalar field, is thus proportional to $1 - \gamma^{\textrm{PPN}}= 1/(2+\omega_0)$. 
 The same proportionality can be found by calculating the momentum transfer to a relativistic test particle passing a massive point-like source to second order in the velocity of the test particle based on the PPN metric corresponding to massless scalar tensor theories obtained, for example, in \cite{nordtvedt_post-newtonian_1970,will_theoretical_1971,will2014confrontation}.
}.
Hence, bounds  on $\omega_0$ could be derived from measurements of $\Delta \delta p_T$ with moving sources. 

In contrast, for $v=0$, scalar-tensor modifications of GR cannot be probed in the limit $m_{\psi_1}\to 0$ as the modification does not anymore depend on the distance between the interacting masses.

\subsection{Particle beams at LHC}\label{sec:LHC}

Lab-based precision experiments to test scalar-tensor theory against GR may be realized at the LHC or next generation accelerator infrastructures with near future high-precision acceleration sensors. At the LHC, 2808 bunches of $1.2\times 10^{11}$ protons with total energy of $E=6.5\,$TeV move in a circular beamline of $27\,$km length. The large circumference implies that the linear propagation model above is a good approximation for realistic acceleration sensors that are usually at most of meter size. Furthermore, the kinetic energy of the proton bunches is well in the ultra-relativistic regime as the proton rest-mass energy is about $938\,$MeV, which leads to $\gamma\sim 7\times 10^3$.  In turn, this implies that the second term in \eqref{eq:relmomentumtransmassive_main} is strongly suppressed in comparison to the first term and the relative momentum transfer difference is equivalent to the expression in equation \eqref{eq:relmomentumtransmassivelim}. To leading order, the  transverse momentum transfer of the bunches will add up linearly to a net effect on the sensor. Therefore,  per passing bunch, we expect a momentum transfer of $\delta p_T\sim M\times 10^{-28}\,\mathrm{m/s}$ on a detector of mass $M$ at a distance of $1\,$cm from the beam line. This effect has to be measured with a relative precision of $10^{-3}$ to give new bounds on the parameters of the scalar-tensor theory in the hitherto unexplored parameter regime considered in figure \ref{fig:rel-plot}, that is, $1/(3+2\omega_0)=10^{-3}$ and $1/m_{\psi_1}=10^2\,$m.

Due to the bunching of the protons, there will be a natural variation of the momentum transfer with bunch periodicity which corresponds to a frequency of $31.2\,$MHz. Another variation can be obtained by filling only half of the accelerator ring leading to a frequency of $11\,$kHz. To achieve lower frequencies, one could, for instance, vary the beam position. If precisely controlled, the variations in the momentum transfer may be used to increase the measurement precision. For a detailed analysis and proposals for realistic detector see \cite{Spengler:2021rlg} by two of us.

It should be noted that the present analysis is performed for neutral source particles while the protons at the LHC are charged. The charge of the source leads to a small correction of the corresponding energy momentum tensor and an additional electromagnetic field acting on the acceleration sensor. The correction to the energy momentum tensor may be investigated in subsequent work. Here, we would like to shortly discuss the electromagnetic field of the source and how to deal with it in a potential experiment. Although the sensor can be assumed to be virtually neutral and electromagnetic forces only arise due to its multipole moments and remaining surface charges, these forces can still be many orders of magnitude larger than the gravitational force on the sensor. Furthermore, since the electromagnetic force on the sensor will oscillate at the same frequency as the gravitational force, it cannot be distinguished through its time dependence and will drive a resonant sensor. Therefore, the sensor has to be shielded from the electromagentic field of the source particles, for example, by enclosing it by a shield of an appropriate material. The properties of the shield (thickness, material etc.) have to be chosen such that the remaining strength of the electromagnetic field on the detector's side is smaller or at least of similar order of magnitude as the modification of the gravitational effect that is to be detected. For example, a metal shield would lead to an exponential suppression of the time-averaged electric field of the source on the length scale of the Thomas-Fermi length which is of the order of one \r{A}  \cite{Ashcroft-Mermin}. The oscillating part of the electric field would be exponentially suppressed on the length scale of the skin depth \cite{brandl2017towards}. Similarly, constant magnetic fields are suppressed exponentially by superconductors on the length scale of the London penetration depth which typically ranges from tens to hundreds of nano-meters \cite{Kittel} and oscillating magnetic fields are suppressed exponentially with the skin depth in the same way oscillating electric fiedls. The ratio of the gravitational force and the remaining electromagnetic forces on the sensor will then depend on the composition and the mass of the chosen sensor. A simple model for the case of a sensor mass of silicon dioxide at a distance of $1\,$cm from the beam line that measures the gravitational field of a transverse distance modulated beam at $100\,$Hz is given in Appendix~\ref{app:electromagnetic}. We find that a shield of a few millimeters thickness suppresses the dominant electric force on uncharged dielectrics, dielectrophoresis, down to one order of magnitude below the gravitational force of the beam. We also conclude that such a shield should be sufficient to suppress all magnetic effects and the effect of charges on the surface of the sensor.  A more detailed analysis will be part of a future investigation in collaboration with researchers that are more familiar with the experimental conditions at the LHC.

\section{Conclusions}\label{sec:conc}

Astrophysical tests of theories of gravity often face the problem, that the conditions of the observation rely on  parameters that cannot be well controlled or influenced by the observer, in contrast to laboratory tests on Earth. Conventionally laboratory tests of gravity are performed with slowly moving source masses, hence testing the Newtonian regime. At the LHC ultrarelativistic proton bunches of well defined energy source a gravitational field which can potentially measured by high-precision acceleration sensors. We showed how such acceleration measurements might be used to distinguish GR and large classes of its scalar-tensor extensions, provided that the sensitivity of the accelerometers employed can be made good enough.

Our main finding is that the parameters of scalar-tensor gravity affect the magnitude and the dependence on source velocity and spatial distance of the acceleration of test masses. The relative momentum transfer difference shows a clear velocity dependence as displayed in equation \eqref{eq:relmomentumtransmassive_main}, whose quantitative behaviour is shown in Fig. \ref{fig:rel-plot}. The figure demonstrates that the deviation from GR can be maximized by increasing the velocity of the source of the gravitational field. Thus, by the measurement of the gravitational field of ultrarelativistic test particles, scalar-tensor theories can be constrained, in a regime which is, to our knowledge, so far not investigated in laboratory experiments. 

Besides scalar-tensor theories, our findings motivate to perform similar studies for general modified theories of gravity, such as Horndeski-Gravity theories beyond the ones we discussed here \cite{Horndeski:1974wa}, metric-affine gravity \cite{Hehl:1994ue}, teleparallel or symmetric teleparallel gravity \cite{Nester:1998mp,Bahamonde:2021gfp,Heisenberg:2023lru} and many more, which for example are nicely displayed in the map of modified gravity in \cite{CANTATA:2021ktz}. The magnitude of the deviation of the test particle acceleration from its GR value, may depend strongly on the theory under investigation, in particular if features like extended spacetime geometries or screening mechanisms are included. This, and the prospect of establishing ultraprecise gravity sensors at the LHC or its successors, makes a systematic
investigation of the coupling of test particles in the different theories in their ultrarelativstic limit highly interesting and worth studying.

\section*{Acknowledgements}
CP acknowledges the financial support by the excellence cluster QuantumFrontiers of the German Research Foundation (Deutsche Forschungsgemeinschaft, DFG) under Germany's Excellence Strategy -- EXC-2123 QuantumFrontiers -- 390837967 and was funded by the Deutsche Forschungsgemeinschaft (DFG, German Research Foundation) - Project Number 420243324. D.R. acknowledges support by the Deutsche Forschungsgemeinschaft (DFG, German Research Foundation) under Germany’s Excellence Strategy – EXC-2123 QuantumFrontiers – 390837967 and the CRC TerraQ from the Deutsche Forschungsgemeinschaft (DFG, German Research Foundation) – Project-ID 434617780 – SFB 1464. DB acknowledges support by the EU EIC Pathfinder project QuCoM (101046973).

\onecolumngrid

\appendix
\vspace{1cm}

\section{Solving the linearized field equations}\label{app:fieldEQ}
We display some details of how to solve the field equations \eqref{eq:metricwaveeq} and \eqref{eq:scalarwaveeq} here. It amounts to solving a massless and a massive wave equation.

A solution of the massive wave equation of the scalar field
\begin{align}
    \eta^{\alpha\beta}\partial_{\alpha }\partial_{\beta}\tilde{\psi}_{1} - m_{\psi_1}^2 \tilde{\psi}_{1}{}  &= \frac{\kappa^2 \tilde{T}}{3+2\omega_0}
\end{align}
for a source at rest ($\tilde{T} = - M c^2 \delta(x) \delta(y) \delta(z)$) is given by the famous Yukawa potential, see for example \cite{Hohmann:2013rba},
\begin{align}
    \tilde \psi_1(\vec x) = \frac{\kappa^2}{4\pi (3+2\omega_0)}\frac{M c^2}{\sqrt{z^2+\rho^2}}e^{-m_\psi\sqrt{z^2+\rho^2}}\,.
\end{align}
As the energy momentum tensor for the moving source is obtained by performing the Lorentz transformation 
\begin{align}
    \Lambda = 
    \begin{pmatrix}
      \gamma & 0 & 0 & \gamma \frac{v}{c}\\
       0 & 1 & 0 & 0 \\
       0 & 0 & 1 & 0 \\
      \gamma\frac{v}{c} & 0 & 0 & \gamma 
\end{pmatrix}\,,
\end{align}
into the reference system of an observer that moves with velocity $v$ in the negative $z$-direction, and accordingly, $T(\vec{x})=\tilde{T}(\Lambda^{-1}\vec{x})$, we find the corresponding solution of the scalar wave equation by the same transformation
\begin{align}
    \psi_1(\vec x) := \tilde \psi_1 (\Lambda^{-1} \vec x) \,.
\end{align}
obtaining our solutions \eqref{eq:psi1}.

The massless wave equation for the trace-reversed metric in scalar-tensor harmonic gauge
\begin{align}
    \eta^{\alpha\beta}\partial_{\alpha }\partial_{\beta}\bar{h}_{\mu \nu } 
    &= -\frac{2 \kappa^2}{\psi_{0}{}} T_{\mu \nu }
\end{align}
is solved by the methods of retarded Greens function using retarded time $t_{\rm ret} = t - \tfrac{|\vec x - \vec x'|}{c}$, analogously of how one obtains the Li\'enard-Wiechert potential of a moving electric charge in electrodynamics. The solution is
\begin{equation}
\begin{aligned}
    \bar h_{\mu\nu} 
    &=  \frac{2\kappa^2}{\psi_0} M \gamma v_\mu v_\nu \frac{1}{4\pi} \int_{\mathbb R^3} dx' dy' dz' \frac{ \delta(x') \delta(y') \delta(z'- v t_{\rm ret})}{|\vec x - \vec x'|}\\
    &= \frac{2\kappa^2}{4 \pi \psi_0} M \gamma v_\mu v_\nu \frac{\gamma}{\sqrt{\rho^2+\gamma^2(z-vt)}}\,,
\end{aligned}
\end{equation}
where one simply evaluates the integral by the applying the rules for a $\delta$-distribution $\delta(f(z'))$ depending on a non-trivial function $f(z')$.

Our solution \eqref{eq:metrich} is now easily obtained by reverting the gauge fixed metric to the general metric and by inserting the solution of the scalar field. As for the scalar field case, the same result can be reached by boosting the diagonal metric perturbation defined in \cite{Hohmann:2013rba} into the reference system of an observer that moves with velocity $v$ in the negative $z$-direction. Starting from  \cite[Eq. (18)]{Hohmann:2013rba} at linear post-Minkowskian ($\kappa^2$ in the bare gravitational constant) order
\begin{align}
    \tilde g(r) - \eta = \tilde h(r)
    = \begin{pmatrix}
      2 G_\mathrm{eff}(r) U(r) & 0 & 0 & 0\\
       0 & 2 G_\mathrm{eff}(r) \gamma (r) U(r) & 0 & 0 \\
       0 & 0 & 2 G_\mathrm{eff}(r) \gamma (r) U(r) & 0 \\
      0 & 0 & 0 & 2 G_\mathrm{eff}(r) \gamma (r) U(r)
\end{pmatrix}\,,
\end{align} 
and using the leading order post-Newtonian results from [Eq. (19), (25), (29)]\cite{Hohmann:2013rba},
\begin{align}
    U(r) = \frac{\kappa^2}{8\pi} \frac{M}{r}\,,\quad 
    G_\mathrm{eff}(r)= \frac{1}{\psi_0} \left(\frac{ e^{-m_{\psi_1} r}}{2 \omega_0+3}+1\right)\,,\quad
    \gamma(r) = \frac{2 \omega_0+ 3 -e^{-m_{\psi_1} r}}{2 \omega_0 + 3 + e^{-m_{\psi_1} r}}\,,
\end{align}
yields the desired metric perturbation components
\begin{align}
    h_{\mu\nu}(\vec x) = \tilde g_{\sigma\rho}(\Lambda^{-1} \vec x )(\Lambda^{-1})^\rho{}{}_{\mu}(\Lambda^{-1})^\sigma{}_{\nu} -\eta_{\mu\nu} = \tilde h_{\mu\nu}(\Lambda^{-1} \vec x)(\Lambda^{-1})^\rho{}{}_{\mu}(\Lambda^{-1})^\sigma{}_{\nu}\,,
\end{align}
which we obtained directly.

\section{The leading and sub-leading acceleration}\label{app:subleading}
The leading and sub-leading terms in $v_T$ of the spatial acceleration \eqref{eq:spatacc1} contribute to the acceleration of a test mass by the gravitational field of a relativistic particle. Here we display the the derivatives of the first order metric and show how to derive the different components of the acceleration. To first order in $v_T$, it is in general given by (see for example \cite[Eq. (9.1.2)]{Weinberg})
\begin{equation}
    \begin{aligned}\label{eq:spatacc}
    \frac{d^2 x_T^i}{dt^2} 
    & = a_T^i = - c^2 \left( \Gamma^i{}_{00} + \left(2  \Gamma^i{}_{j0} - \Gamma^0{}_{00} \delta_j^i\right) \frac{v_T^j}{c}\right)\\
    &= \frac{c}{2}\eta^{ij} \big(c\ \partial_j h_{00} - 2 \partial_t h_{j0}   
    - v_T^k (2\partial_k h_{0j}-2\partial_j h_{0k}   + \tfrac{2}{c} \partial_t  h_{jk}  + \tfrac{1}{c} \eta_{jk}  \partial_t h_{00})  \big)\,.
    \end{aligned}
\end{equation}
In order to derive the magnitude of the spatial acceleration we need the derivatives of the metric components \eqref{eq:metrich}, which are most conveniently expressed in terms of the derivatives acting on the function $\mathcal{H}$ and $\mathcal{K}$ (see \eqref{eq:defK} and \eqref{eq:defH}) which in turn can be expressed through $\rho_i$ with $\rho_1 = x_1 = x, \rho_2 = x_2 = y, \rho_3 = 0$:
\begin{align}
    \nonumber \partial_j h_{00} 
    &= \gamma^2 \partial_j \mathcal{K} - \tfrac{1}{2}\partial_j \mathcal{H}\\
    &= -\frac{2 \kappa^2 M c^2}{4 \pi \psi_0} \frac{\left(\rho_j + \gamma^2 (z-vt)\delta_{3j}\right)}{(\rho^2 + \gamma^2 (z - v t)^2)^{3/2}} \left[\gamma^2 - \tfrac{1}{2}  + \tfrac{m_{\psi_1}\sqrt{\rho^2 + \gamma^2 (z - v t)^2}+1}{2(3+2\omega_0)}e^{-m_{\psi_1}\sqrt{\rho^2 + \gamma^2 (z - v t)^2}}  \right]\,,\\
    \partial_t h_{00}
    &= \gamma^2 \partial_t \mathcal{K} - \tfrac{1}{2}\partial_t \mathcal{H}
    \\
    \partial_t h_{j0}
    &= - \tfrac{\gamma^2 v}{c} \delta^3_j \partial_t \mathcal{K}\\
    \partial_j h_{k0}
    &= - \tfrac{\gamma^2 v}{c} \delta^3_k \partial_j \mathcal{K}\\
    \partial_t h_{ij}
    &= \tfrac{\gamma^2 v^2}{c^2} \delta^3_i \delta^3_j \partial_t \mathcal{K} + \tfrac{1}{2}\eta_{ij}\partial_t \mathcal{H}\,.
\end{align}
For the leading order contribution $\mathcal{O}((v_T)^0)$ we see that the only relevant leading order term is the one coming from $\partial_K h_{00} = \gamma^2 \partial_K \mathcal{K}- \partial_K \mathcal{H}/2$, as claimed above \eqref{eq:atrans1}. The terms $\partial_t h_{0j}$ and $\partial_3 h_{00}$ are not relevant, since they can be rewritten as being proportional to $\partial_t \mathcal{K}$ or $\partial_t \mathcal{H}$ by the identities \eqref{eq:dtKtod3K}. When these then get integrated from $t=-\infty$ to $\infty$ in the momentum transfer integral \eqref{eq:deltapscalartensor}, this simply leads to an evaluation of $ \mathcal{K}$ and $ \mathcal{H}$ at $t=-\infty$ and $t=\infty$, where they vanish by the limits \eqref{eq:limK} and \eqref{eq:limH}.

\subsection{Sub-leading transversal acceleration}\label{app:trans}

The sub-leading component of the transverse acceleration, is 
\begin{equation}
    \begin{aligned}\label{eq:atrans2}
    \bar  a_T^{K}
    &= -\frac{c}{2}\eta^{Kj} v_T^i (2\partial_i h_{0j}-2\partial_j h_{0i}   + \tfrac{2}{c} \partial_t  h_{ji}  + \tfrac{1}{c} \eta_{ji}  \partial_t h_{00}) \\
    &= - \gamma^2 v v^3_T \partial^K \mathcal{K}  - \tfrac{1}{2} v_T^K (\gamma^2 \partial_t \mathcal{K} + \tfrac{1}{2}\partial_t \mathcal{H})\,.
    \end{aligned}
\end{equation}
We observe that the only relevant term for the momentum transfer is the first one, since it is the only term with no $t$-derivative acting on $\mathcal{K}$ or $\mathcal{H}$, thus the only term which contributes to the integration over time from $t=-\infty$ to $\infty$ \eqref{eq:deltapscalartensor}. This relevant term is suppressed with respect to the leading order term by a factor $v^3_T/c$.

\subsection{The longitudinal acceleration}\label{app:longi}
At the end of Sec. \ref{ssec:spatacc}, we mentioned that the longitudinal acceleration does not contribute to the momentum transfer to leading order. The argument is the following.

The longitudinal acceleration can be expressed as follows
\begin{equation}
\begin{aligned}
    a_T^3 &=  \frac{c}{2}\eta^{3j} \big(c\ \partial_j h_{00} - 2 \partial_t h_{j0}   
    - v_T^k (2\partial_k h_{0j}-2\partial_j h_{0k}   + \tfrac{2}{c} \partial_t  h_{jk}  + \tfrac{1}{c} \eta_{jk}  \partial_t h_{00})  \big)  \\
    &=  \frac{c^2}{2v} \left( (\gamma^2-2) \partial_t\mathcal{K} + \frac{1}{2}\partial_t \mathcal{H}    \right) + \gamma^2 v v_T^K\partial_K \mathcal{K} - \frac{v_T^3}{2} \left( (3\gamma^2-2) \partial_t\mathcal{K}  +  \frac{1}{2}\partial_t \mathcal{H} \right)
    \,.
\end{aligned}
\end{equation}
As time derivatives do not contribute to the momentum transfer integral, the only contibuting term in the longitudinal acceleration \eqref{eq:deltapscalartensor} is
\begin{align}
    -c v_T^K \partial_K h_{30} = \gamma^2 v v_T^K \partial_K \mathcal{K}\,.
\end{align}
As for the transversal acceleration, this term is suppressed with respect to the general leading order term by a factor of the test particle velocity over the speed of light.

\section{Curvature and Geodesic deviation}\label{app:curvgeoddevi}
In this appendix we display the mathematical details about the curvature and geodesic deviation (relative acceleration of test particles), which are discussed in Sec. \ref{ssec:geoddevi}.

\subsection{The curvature of the linearized metric}\label{app:Curv}

To calculate the components of the linearized curvature, we employ the metric perturbation as in \eqref{eq:metrich}. 
We obtain
\begin{equation}
	\begin{aligned}
    R_{\alpha\beta\gamma\delta} &= \frac{1}{2}\left(\partial_\beta\partial_\gamma h_{\delta\alpha} - \partial_\beta\partial_\delta h_{\gamma\alpha} - \partial_\gamma\partial_\alpha h_{\beta\delta} + \partial_\delta\partial_\alpha h_{\beta\gamma}\right) \\
    &=  \frac{\gamma^2}{2c^2}\left( v_\delta v_\alpha \partial_\beta\partial_\gamma   -  v_\gamma v_\alpha \partial_\beta\partial_\delta   - v_\beta v_\delta  \partial_\gamma\partial_\alpha   + v_\beta v_\gamma  \partial_\delta\partial_\alpha   \right)\mathcal{K} \\
    &\quad + \frac{1}{4}\left(\eta_{\delta\alpha}\partial_\beta\partial_\gamma   - \eta_{\gamma\alpha}   \partial_\beta\partial_\delta   - \eta_{\beta\delta} \partial_\gamma\partial_\alpha   + \eta_{\beta\gamma} \partial_\delta\partial_\alpha   \right) \mathcal{H} \,.
    \end{aligned}
\end{equation}
Using $v_\mu = (-c,0,0,v)$ and the relations \eqref{eq:limH}, 
we find for the components of the curvature tensor with two time-indices 
\begin{align}
    \label{eq:R0303}  R_{0303} &=   - \frac{c^2}{2v^2\gamma^2}\partial_0\partial_0 \left(\mathcal{K} - \frac{1}{2} \mathcal{H}\right)  \\
    \label{eq:R030K}  R_{030K} &=   \frac{c}{2v}\partial_0 \partial_K \left(\mathcal{K} - \frac{1}{2} \mathcal{H}\right)  \\
    R_{033K} &= 
    -\frac{v}{c}  R_{030K}\\
    R_{0K0J} &=  -\frac{1}{2}\partial_K\partial_J \left(\gamma^2 \mathcal{K} - \frac{1}{2} \mathcal{H}\right) - \frac{1}{4}\eta_{KJ} \partial_0\partial_0 \mathcal{H}   \\
    R_{03JK} &=  0 \\
    R_{0J3K} &= \frac{v}{2c}\left(\gamma^2 \partial_J\partial_K \mathcal{K}  + \frac{c^2}{2v^2} \eta_{JK} \partial_0\partial_0 \mathcal{H}  \right) \\
    R_{0JKL} &= 
    = -\frac{v}{c}R_{3JKL} \\
    R_{3JKL} &=  -\frac{c}{4v}\left(\eta_{JK} \partial_L  - \eta_{JL}\partial_K \right) \partial_0 \mathcal{H} \\
    R_{3J3K} &=  -\frac{\gamma^2v^2}{2c^2}\partial_J\partial_K \mathcal{K} -\frac{1}{4}\left(\partial_J\partial_K + \frac{c^2}{v^2}\eta_{JK}\partial_0\partial_0\right)\mathcal{H} \\
    R_{JKLM} &=   \frac{1}{4}\left(\eta_{MJ}\partial_K\partial_L   - \eta_{LJ}\partial_K\partial_M   - \eta_{KM} \partial_L\partial_J   + \eta_{KL} \partial_M\partial_J \right) \mathcal{H} \,,
\end{align}

\subsection{Longitudinal Geodesic deviation}\label{app:geoddevi}
At the end of Sec. \ref{ssec:geoddevi}, we claimed that the longitudinal relative acceleration of a sensor test particle is not relevant for our discussion. We prove this statement here.  

The longitudinal differential acceleration of two test particles at an infinitesimal distance from each other
is given by
\begin{align}
    \ddot s^3 =  R^3{}_{00K}s^K 
    + R^3{}_{003}s^3.
\end{align}
The relevant curvature components are derived in \eqref{eq:R030K} and \eqref{eq:R0303}. Since there is no curvature at zeroth order, there is no accelerated geodesic deviation at zeroth order and we can write $s^\mu(t) = s^\mu_0 + v^\mu_0 t + \Delta s^\mu(t)$, for some constants $s_0^\mu$ and $v^\mu_0$. In addition we assume that the initial geodesics are at rest with respect to each other, which means that the constant which parametrizes the initial relative velocity of nearby geodesics vanishes $\vec{v}_0 = 0$, not to be confused with the velocity of the test sensor $\vec{v}_T$ or the velocity of the particle beam $\vec{v}$. The curve parameter is chosen here to be the Minkowski spacetime coordinate time $t$, which is possible since at zeroth order ($h_{\mu\nu}\to 0$), the Minkowski spacetime coordinates can be chosen such that $t$ is precisely the arc-length parameter of geodesics at rest with respect to each other.

The first non-trivial order $\Delta s^\mu(t)$ which are sourced by the non-trivial gravitational field produced by the particle beam are of the same order as the metric and the scalar field perturbation, hence of the same order as the curvature.

The integration of the geodesic deviation from $-\infty$ to $\infty$ over time then becomes 
\begin{equation}
    \begin{aligned}
    \int^{\infty}_{-\infty} dt\ \ddot  s^3
    \approx \int^{\infty}_{-\infty} dt\,  \frac{d^2}{dt^2}\Delta s^3 
    &=  \int^{\infty}_{-\infty} dt\, s_0^K R^3{}_{00K} +  \int^{\infty}_{-\infty} dt\, s_0^3 R^3{}_{003}\\\
    &= - \int^{\infty}_{-\infty} dt\, (s_0^K R_{030K} +  s_0^3 R_{0303} )\\
    &= -\int^{\infty}_{-\infty} dt \frac{c}{2v} \partial_0 \left( (s_0^K \partial_K -  \tfrac{1}{\gamma^2 }\tfrac{c}{v} s_0^3 \partial_0)(\mathcal{K} - \tfrac{1}{2}\mathcal{H}) \right)\\
    &= -\int^{\infty}_{-\infty} dt \frac{1}{2v} \partial_t \left( (s_0^K \partial_K -  \tfrac{1}{\gamma^2 }\tfrac{1}{v} s_0^3 \partial_t)(\mathcal{K} - \tfrac{1}{2}\mathcal{H}) \right)\\
    &= \frac{1}{2 v} \left[  (s_0^K \partial_K -  \tfrac{1}{\gamma^2 }\tfrac{1}{v} s_0^3 \partial_t)
    (\mathcal{K} - \tfrac{1}{2}\mathcal{H})\right]_{t=-\infty}^{t = \infty}=0\,,
    \end{aligned}
\end{equation}
due to the identities \eqref{eq:limK} and \eqref{eq:limH}, which proves our original claim. Physically this result means that the longitudinal acceleration caused by the gravitational field of the particle beam is not detectable as long as the reaction of the sensor is slow compared to the time scale on which the particle passes by.

\section{Calculation of momentum transfer}

In this section, we present a step-by-step derivation of the momentum transfer in equation \eqref{eq:deltapscalartensor}. From equation \eqref{eq:atrans1}, we find
\begin{equation}
    \begin{aligned}
         \delta p^K &\approx - \frac{\kappa^2 M m c^4}{4\pi\psi_0} \int_{-\infty}^\infty dt \,\frac{x^K}{(\gamma^2(z-vt)^2+\rho^2)^{3/2}} \left( \gamma^2  - \frac{1}{2}  + \frac{1 + m_\psi\sqrt{\gamma^2(z-vt)^2+\rho^2}}{2(3+2\omega_0)}  e^{-m_{\psi_1}\sqrt{\gamma^2(z-vt)^2+\rho^2}}\right)   \\
        &= -\frac{\kappa^2 M m c^4}{4\pi\psi_0}\frac{ x^K}{\gamma v} \int_{-\infty}^\infty d\xi \,\left(\gamma^2 - \frac{1}{2} + \frac{1 + m_\psi\sqrt{\xi^2+\rho^2}}{2(3+2\omega_0)}  e^{-m_{\psi_1}\sqrt{\xi^2+\rho^2}}\right)\frac{1}{(\xi^2+\rho^2)^{3/2}} \\
        &= -\frac{\kappa^2 M m c^4}{4\pi\psi_0}\frac{ x^K}{\gamma v} \left( \left(\gamma^2 - \frac{1}{2}\right)\frac{2}{\rho^2} -\frac{1}{2(3+2\omega_0)}\frac{1}{\rho} \frac{d}{d\rho} \int_{-\infty}^\infty d\xi \,\frac{e^{-m_{\psi_1}\sqrt{\xi^2+\rho^2}}}{\sqrt{\xi^2+\rho^2}} \right)\\ 
        &= -\frac{\kappa^2 M m c^4}{4\pi\psi_0}\frac{ x^K}{\gamma v} \left( \left(\gamma^2 - \frac{1}{2}\right)\frac{2}{\rho^2} -\frac{1}{2(3+2\omega_0)}\frac{1}{\rho} \frac{d}{d\rho}  2\int_{\rho}^\infty d\sigma \,\frac{e^{-m_{\psi_1}\sigma }}{\sqrt{\sigma^2-\rho^2}}  \right)\\
        &= -\frac{\kappa^2 M m c^4}{4\pi\psi_0}\frac{ x^K}{\gamma v} \left( \left(\gamma^2 - \frac{1}{2}\right)\frac{2}{\rho^2} -\frac{1}{2(3+2\omega_0)}\frac{1}{\rho} \frac{d}{d\rho}  2 K_0(m_{\psi_1}\rho) \right)\\
        &= -\frac{\kappa^2 M m c^4}{4\pi\psi_0}\frac{ x^K}{\gamma v} \left( \left(\gamma^2 - \frac{1}{2}\right)\frac{2}{\rho^2} +\frac{1}{(3+2\omega_0)}\frac{m_{\psi_1}}{\rho}    K_1(m_{\psi_1}\rho) \right)
    \end{aligned}
\end{equation}
where $K_0$ and $K_1$ are modified Bessel functions of the second kind.

\section{Electromagnetic forces on the sensor}\label{app:electromagnetic}

In the following, we will derive an upper bound for the ratio of the electromagnetic force and the gravitational force acting on a dielectric force sensor in the vicinity of the LHC beam, to estimate the thickness of the necessary shielding (most likely in form of a Faraday cage) to measure the pure gravitational effect, as discussed in section \ref{sec:LHC}. 

The setup we consider is a gravitational field sensor, a test mass, at an average distance $\rho$ to the particle beam.  The particle beam is slightly modulated back and forth by a small amount $\delta\rho$ at a frequency $\Omega$ which we assume to be about 100\,Hz \cite{Spengler:2021rlg}. Ideally, the sensor will be enclosed in a shield of metal, i.e.~a Faraday cage, to reduce the electromagnetic effect. For a simple estimate of the remaining electric field due to the charged protons that is affecting the sensor at its position, we consider the shielding with an infinitely extended metal shield.

The sensor is assumed to have a narrow resonance which is tuned to the modulation frequency.

For the purpose of our estimates in this appendix, it is sufficient to model the proton beam as an infinitely long cylindrical charge distribution along the $z$-direction. In its rest frame, the magnetic field vanishes and the electric field takes the form
\begin{equation}
    \boldsymbol{E}=\frac{\lambda}{2\pi\varepsilon_0\rho}\hat{\rho}\,,
\end{equation}
where $\hat{\rho}$ is the unit vector perpendicular to the beam and $\lambda$ is the line charge density. A Lorentz transformation into the lab frame (the rest frame of the shield and the sensor) leads to the expressions for electric field and magnetic field
\begin{align}
    \boldsymbol{E}&=\frac{j}{2\pi\varepsilon_0 v\rho}\hat{\rho} \quad\text{and}\quad
    \boldsymbol{B} =\frac{\mu_0 j}{2\pi \rho}\hat{z}\times\hat{\rho}
\end{align}
where $j=v\gamma\lambda$ is the beam current in the lab frame given by $e$ times the number $N$ of protons that pass the detector per second leading to $j=e\nu N$, where $\nu$ is the circulation frequency. 
We assume that $\rho$ is modulated by $\delta\rho\ll\rho$ in a monochromatic fashion and slowly enough (negligible retardation) such that, to leading order, electric and magnetic field can be written as $\boldsymbol{E}_\mathrm{mod}=\boldsymbol{E}+\delta\boldsymbol{E}\cos(\Omega t+\varphi)$ and $\boldsymbol{B}_\mathrm{mod}=\boldsymbol{B}+\delta\boldsymbol{B}\cos(\Omega t+\varphi)$, respectively, where the modulation amplitude vectors at the resonance frequency of the sensor are
\begin{align}
    \delta\boldsymbol{E} &=\frac{j\delta\rho}{2\pi\varepsilon_0 v\rho^2} \hat{\rho}\quad\text{and}\quad
    \delta\boldsymbol{B} =\frac{\mu_0 j\delta\rho}{2\pi \rho^2}\hat{z}\times\hat{\rho}\,.
\end{align}
For the gravitational force on the sensor due to the beam, we can use the approximate expression \cite{Spengler:2021rlg}
\begin{equation}
    \boldsymbol{F}_g = - \frac{4G m P}{c^3\rho}\hat{\rho}\,,
\end{equation}
where $G$ is Newton's gravitational constant, $m$ is the sensor probe mass and $P$ is the circulating beam power that is related to the current as $P=E_\mathrm{p} j/e$ and $E_\mathrm{p}=\gamma M c^2$ is the proton energy. The amplitude of the gravitational force oscillations at the sensor's resonance frequency is then
\begin{equation}
    \delta\boldsymbol{F}_g = \frac{4GmP\delta\rho}{c^3\rho^2}\hat{\rho}\,.
\end{equation}

While the gravitational force field passes the shield virtually unchanged, the electromagnetic field is strongly suppressed. In the quasistatic regime, the suppression of the electric field and magnetic field strength are exponential on length scale of the Thomas-Fermi length $l_\mathrm{F}$ \cite{Ashcroft-Mermin}
and the London penetration depth 
$l_\mathrm{L}$ \cite{Kittel}, respectively. Therefore, we obtain behind the shield of thickness $d$
\begin{align}
    \boldsymbol{E}_\mathrm{s} &=  \boldsymbol{E} e^{-d/l_\mathrm{F}}\quad\text{and}\quad
    \boldsymbol{B}_\mathrm{s} = \boldsymbol{B} e^{-d/l_\mathrm{L}} \,,
\end{align}
For the modulations $\delta\boldsymbol{E}$ and $\delta\boldsymbol{B}$ the screening length \begin{equation}
    l_\text{s}=\sqrt{\frac{2}{\omega \sigma \mu}}\,,
\end{equation}
where $\sigma$ is the conductivity and $\mu$ the permeability of the metal, is the relevant length scale for the exponential damping. 
In copper at room temperature, the skin depth at a frequency of $2\pi\times 50$\,Hz is about 9.2\,mm \cite{brandl2017towards}. Since the conductivity increases exponentially when lowering the temperature, one can very effectively screen electromagnetic fields.  At liquid helium temperature, the conductivity of copper increases by a factor of about 1000 compared to room temperature (see Fig. 4.6 in \cite{brandl2017towards}). Therefore, the skin depth for 100 Hz signals in copper at 4.2\,K  is expected to be of the order of 0.2\,mm.
The Thomas-Fermi length is usually of the length scale of \r{A} \cite{Ashcroft-Mermin} and the London penetration depth can be assumed to be of the order of tens to hundreds of nanometers (e.g.~Lead $\lambda(0)\le 30.5$\,nm \cite{PhysRevB.2.2519}, Niobium $\lambda(0)=47\pm5$\,nm \cite{PhysRev.139.A1515} with temperature dependence $\lambda(T)=\lambda(0)\left(1-\left(T/T_C\right)^4\right)^{-1 / 2}$). In the following, we will make the conservative assumption that $l_\mathrm{F}\sim 1$~nm and $l_\mathrm{L}\sim 100$~nm.

Electric and magnetic field behind the shield are acting on a dielectric sensor that is uncharged up to a small number of remaining surface charges. We will discuss the surface charges below and focus on the dielectrophoretic force in the following. For a conservative estimate, we model the sensor mass as a small (almost pointlike) lossless (vanishing conductivity) dielectric sphere of radius $R$ for which the dielectrophoretic force is \cite{jones1995electromechanics}
\begin{equation}
    \boldsymbol{F}_\mathrm{ED}= 2\pi \varepsilon_1 R^3 K_\mathrm{CM} \nabla |\boldsymbol{E}_\mathrm{s} + \delta\boldsymbol{E}_\mathrm{s}\cos(\Omega t + \varphi) |^2 \,,
\end{equation}
where $K_\mathrm{CM}=(\epsilon_1-\epsilon_0)/(\epsilon_1+2\epsilon_0)$ enters through the Clausius-Mossotti relation and $\varepsilon_1$ is the permittivity of the dielectric material. The force amplitude vector at the resonance frequency of the sensor becomes 
\begin{equation}
    \delta\boldsymbol{F}_\mathrm{ED,1}= 2\pi \varepsilon_1 R^3 K_\mathrm{CM} \nabla (2\delta\boldsymbol{E}_\mathrm{s}\cdot\boldsymbol{E}_\mathrm{s}) = -12\pi \varepsilon_1 R^3 K_\mathrm{CM} \left(\frac{j}{2\pi\varepsilon_0 v\rho^2}\right)^2  \delta\rho \, e^{-d/l_\mathrm{F}-d/l_\text{s}}  \hat{\rho} \,.
\end{equation}
Additionally, there appears a second oscillating term in the dielectrophoretic force
\begin{equation}
    \delta\boldsymbol{F}_\mathrm{ED,2}= 2\pi \varepsilon_1 R^3 K_\mathrm{CM} \nabla (\delta\boldsymbol{E}_\mathrm{s}\cdot\delta\boldsymbol{E}_\mathrm{s}) = -8\pi \varepsilon_1 R^3 K_\mathrm{CM} \left(\frac{j}{2\pi\varepsilon_0 v\rho^2}\right)^2 \frac{\delta\rho^2}{\rho}  \, e^{-2d/l_\text{s}}  \hat{\rho} \,.
\end{equation}
While $\delta\boldsymbol{F}_\mathrm{ED,2}$ may be larger than $\delta\boldsymbol{F}_\mathrm{ED,1}$, it oscillates at twice the modulation frequency which implies that it is strongly suppressed if the resonance of the sensor 
or the bandwidth of its readout 
is chosen narrow enough. 
The quotient of the amplitude of dielectrophoretic force $\delta\boldsymbol{F}_\mathrm{ED,1}$ and gravitational force is then
\begin{align}
    \frac{|\delta\boldsymbol{F}_\mathrm{ED,1}|}{|\delta\boldsymbol{F}_g|} &= 12\pi \varepsilon_1 R^3 K_\mathrm{CM}   \left(\frac{j}{2\pi\varepsilon_0 v\rho^2}\right)^2\left(\frac{4Gm E_\mathrm{p} j}{c^3e\rho^2}\right)^{-1} e^{-d/l_\mathrm{F}-d/l_\text{s}}   \\
    & = 9 \varepsilon_1 K_\mathrm{CM}    \left(\frac{j}{2\pi\varepsilon_0 v\rho^2}\right)^2\left(\frac{4G \varrho_m E_\mathrm{p} j}{c^3e\rho^2}\right)^{-1} e^{-d/l_\mathrm{F}-d/l_\text{s}} \\
    & = \frac{9}{16\pi^2} \frac{\varepsilon_1}{\varepsilon_0 }  K_\mathrm{CM} \frac{e^2  N_\mathrm{p}}{\varepsilon_0 G \varrho_m \rho^2 \gamma M l_\mathrm{r} } \frac{c}{v} e^{-d/l_\mathrm{F}-d/l_\text{s}}   \,,
\end{align}
where $l_\mathrm{r}$ is the length of the accelerator ring, $N_\mathrm{p}$ is the number of protons in the ring and $\varrho_m$ is the mass density of the sensor probe. For silicon dioxide, we have $\varepsilon_1=3.9\varepsilon_0$ and $\varepsilon_1(\varepsilon_1-\epsilon_0)/(\varepsilon_0(\varepsilon_1+2\epsilon_0))\approx 2$. Furthermore, for ultra-relativistic particles, we can approximate $v/c=1$ and find the approximate result
\begin{align}
    \frac{|\delta\boldsymbol{F}_\mathrm{ED,1}|}{|\delta\boldsymbol{F}_g|}
    & \approx \frac{9}{8\pi^2} \frac{e^2  N_\mathrm{p}}{\varepsilon_0 G \varrho_m \rho^2 \gamma M l_\mathrm{r} } e^{-d/l_\mathrm{F}-d/l_\text{s}}    \,.
\end{align}
If we consider the values for the LHC, that is, $l_r=26 659\,$m, $N_\mathrm{p}=2.8\times 10^{14}$ protons in the ring (2800 bunches with $10^{11}$ protons each), and $\gamma=7460$ (total energy of 7~TeV), and if we assume a distance of the sensor from the beam line of $1\,$cm, we find for the factor in front of the exponential function a value of about $2\times 10^{16}$. For $l_\mathrm{F}\sim 1\,$nm and $l_\mathrm{s}\sim 0.2\,$mm, a shield of about $40\,$nm thickness would be sufficient to make the dielectrophoretic force term $\delta\boldsymbol{F}_\mathrm{ED,1}$ smaller than the gravitational force by one order of magnitude. Note that, for the chosen parameters, this damping is entirely due to the strong suppression of the constant term in the electric field strength with the Thomas-Fermi length as $d\ll l_\mathrm{s}$. The estimate of $40\,$nm may be a bit too optimistic, however, we conclude that a shield thickness of the order of a millimeter should be more than sufficient for the suppression of $\delta\boldsymbol{F}_\mathrm{ED,1}$.

For the term in the dielectrophoretic force that is quadratic in the modulation of the electric field strength, we find the ratio
\begin{align}
    \frac{|\delta\boldsymbol{F}_\mathrm{ED,2}|}{|\delta\boldsymbol{F}_g|}
    & \approx \frac{3}{4\pi^2} \frac{e^2  N_\mathrm{p}}{\varepsilon_0 G \varrho_m \rho^2 \gamma M l_\mathrm{r} } \frac{\delta\rho}{\rho} e^{-2d/l_\text{s}}    \,.
\end{align}
The $\delta\boldsymbol{F}_\mathrm{ED,2}$ oscillates at twice the resonance frequency of the sensor and the effect on the sensor will be further suppressed. Assuming a Lorentzian response function of the sensor, at driving frequency $\omega$
the response of the detector
is proportional to $\Gamma/\sqrt{(\omega-\Omega)^2+\Gamma^2}$ with resonance width $\Gamma$ (FWHM of the power). The resonance width is related to the quality factor as $Q=\Omega/\Gamma$. Assuming $Q\sim 10^9$, we obtain a damping at $\omega=2\Omega$ of the order of $Q^{-1}/\sqrt{1+1/Q^2}\approx Q^{-1} \sim 10^{-9}$. This implies that it would be sufficient to achieve a damping of $\delta\boldsymbol{F}_\mathrm{ED,2}$ by 7 orders of magnitude through the shielding which corresponds to a thickness $d=1.6\,$mm for $l_\text{s}\approx 0.2\,$mm for copper at 4.2\,K.

The static part of the magnetic field decays 
on the length scale $l_\mathrm{L}$, which we assume to be two orders of magnitude larger than $l_\mathrm{F}$, and the oscillating part decays on the scale of the skin depth. Furthermore, the sensor is only diamagnetic, and therefore, the magnetic force will be significantly smaller than the dielectrophoretic force. Therefore, we conclude that a shield thickness of the order of a few millimeters should be sufficient to suppress the magnetic force well below the gravitational force on the sensor.

We can estimate the number of allowed surface charges on the sensor by calculating the Coulomb force per surface charge on the sensor behind the shield. To simplify the calculations, we neglect the effect of the polarizable material of the sensor which would change electric field strength acting on the charge. 
Then, the Coulomb force is
$\delta\boldsymbol{F}_\mathrm{sf}=e\delta\boldsymbol{E}_\mathrm{s}$ and the quotient with the gravitational force
\begin{align}
    \frac{|\delta\boldsymbol{F}_\mathrm{sf}|}{|\delta\boldsymbol{F}_g|} &=  \frac{ej\delta\rho}{2\pi\varepsilon_0 v\rho^2} \left( \frac{4GmE_\mathrm{p} j\delta\rho}{c^3e\rho^2}\right)^{-1} e^{-d/l_\text{s}} = \frac{e^2}{8\pi\varepsilon_0 Gm \gamma M } \frac{c}{v}e^{-d/l_\text{s}} \approx \frac{e^2}{8\pi\varepsilon_0 Gm \gamma M } e^{-d/l_\text{s}} \,.
\end{align}
If we assume $m=100\,$mg for the sensor mass, we find a value of $10^9$ for the factor in front of the exponential function. A shield of $5\,$mm thickness, would lead to an allowed number of surface charges of the order of $10^{6}$.  Single charge detection on suspended nano-mechanical sensors has recently been demonstrated \cite{PhysRevLett.133.023602}.  This shows that also this effect can be effectively shielded.

In summary, we showed that, in principle, it is possible to shield a sensor in the vicinity of a beam of ultra-relativistic particles well enough that the electromagnetic force signals experienced by a resonant sensor are smaller than the gravitational signal.

\bibliography{PB}

\end{document}